\documentclass[journal,comsoc]{IEEEtran}

\usepackage[T1]{fontenc}
\usepackage{graphicx}
\usepackage{multicol}
\usepackage{multirow}
\usepackage{amsmath,amsfonts,amssymb}
\usepackage{graphicx}
\usepackage{wrapfig}
\usepackage{float}
\usepackage{enumerate}
\usepackage{url}
\usepackage{multirow}
\usepackage{wrapfig}
\usepackage{lipsum}
\usepackage{color}
\usepackage[colorlinks=true, allcolors=blue]{hyperref}
\usepackage{hhline} %
\usepackage{stfloats}

\usepackage[normalem]{ulem}
\newcommand{\umbehide}[1]{}

\ifCLASSINFOpdf
\else
\fi
\usepackage{amsmath}
\usepackage{amsmath}
\usepackage{cite}
\usepackage{float}
\usepackage{placeins}
\usepackage[cmintegrals]{newtxmath}
\usepackage{fixltx2e}
\begin{document}

\title{3-D Stochastic Numerical Breast Phantoms for Enabling Virtual Imaging Trials of Ultrasound Computed Tomography}

%
%
%

\author{Fu~Li,~\IEEEmembership{Student Member,~IEEE,}
        Umberto~Villa,~\IEEEmembership{Member~IEEE,}
        Seonyeong~Park,~\IEEEmembership{Member~IEEE,}
        and~Mark~A.~Anastasio,~\IEEEmembership{Senior Member,~IEEE}
\thanks{F. Li, S. Park, and M. Anastasio are with the Department
of Bioengineering, University of Illinois at Urbana-Champaign, Urbana,
IL, 61801 USA e-mail: \url{maa@illinois.edu}.}
\thanks{U. Villa is with the Electrical \& Systems Engineering Department, Washington University in St. Louis, St Louis, MO, 63130 USA.}
\thanks{Manuscript received \today; revised ---.}}

%
%

\markboth{IEEE Transactions on Ultrasonics, Ferroelectrics, and Frequency Control
Submit Manuscript}%
{Li \MakeLowercase{\textit{et al.}}: A computer-simulation framework for breast ultrasound computed tomography that employs anatomically realistic numerical phantoms}
%



 \maketitle

\begin{abstract}
Ultrasound computed tomography (USCT) is an emerging imaging modality for breast imaging that can produce quantitative images that depict the acoustic properties of tissues. Computer-simulation studies, also known as virtual imaging trials, provide researchers with an economical and convenient route to systematically explore imaging system designs and image reconstruction methods. 
When simulating an imaging technology intended for clinical use, it is essential to employ realistic numerical phantoms that can facilitate the objective, or task-based, assessment of image quality. Moreover, when computing objective image quality measures, an ensemble of such phantoms should be employed that display the variability in anatomy and object properties that is representative of the to-be-imaged patient cohort. Such stochastic phantoms for clinically relevant applications of USCT are currently lacking.
In this work, a methodology for producing realistic three-dimensional (3D) numerical breast phantoms for enabling clinically-relevant computer-simulation studies of USCT breast imaging is presented. By extending and adapting an existing stochastic 3D breast phantom for use with USCT, methods for creating ensembles of numerical acoustic breast phantoms are established. These breast phantoms will possess clinically relevant variations in breast size, composition, acoustic properties, tumor locations, and tissue textures.
To demonstrate the use of the phantoms in virtual USCT studies, two brief case studies are presented that address the development and assessment of image reconstruction procedures. 
 Examples of breast phantoms produced by use of the proposed methods and a collection of 52 sets of simulated USCT measurement data have been made open source for use in image reconstruction development.
\end{abstract}

%
\IEEEpeerreviewmaketitle

\section{Introduction}
%
%
%
%
\IEEEPARstart{U}{ltrasound} Computed Tomography (USCT) is an  imaging technique that utilizes tomographic principles to obtain quantitative estimates of acoustic properties such as speed-of-sound (SOS), density, and acoustic attenuation (AA) \cite{pratt2007sound,wang2015waveform,schreiman1984ultrasound,li2009vivo,duric2007detection}.
Because it can produce high resolution and high contrast images of tissue properties, the development of USCT as a breast imaging modality has recieved significant attention \cite{duric2018breast,nam2013quantitative,greenleaf1977quantitative,duric2007detection,sandhu2015frequency,malik2016objective}. 
It has several advantages over other breast imaging modalities, such as mammography, including low cost and being radiation- and breast-compression-free\cite{ruiter20123d,duric2013breast}.
While commercial systems for breast USCT are being actively developed, USCT remains an emerging technology and a 
topic of active research \cite{taskin20203d,duric2020using,wiskin2020full,javaherian2020refraction}.

When developing new breast USCT technologies, it is important to assess their clinical utility by use of objective measures of image quality (IQ).
Given the large number of system parameters that can impact image quality and variability in the cohort of subjects to-be-imaged, a comprehensive assessment and refinement of modern imaging technologies such as breast USCT via clinical trials often is impossible. Furthermore, obvious ethical concerns preclude certain experimental designs that otherwise would be of great benefit toward optimizing imaging systems for diagnostic tasks, such as tumor detection and characterization. As a surrogate for clinical trials, computer-simulation studies of medical imaging technologies, also known as \emph{virtual imaging trials} (VITs), have been advocated for assessing and optimizing system and algorithm designs\cite{badano2018evaluation,badano2021silico,abadi2020virtual,park2020realistic}. VITs provide a convenient, safe and cost-effective way to explore system and algorithm designs in the early stages of technology development \cite{samei2020virtual,maidment2014virtual}.


For use in computing objective, or task-based, measures of IQ that serve as figures-of-merit (FOM) for breast USCT designs, it is critical that VITs employ numerical breast phantoms (NBPs) that accurately convey the anatomical and acoustic properties of the female breast. 
Moreover, it is known that object variability (i.e., patient-to-patient differences in the breast anatomy and properties) can be viewed as a source of randomness present in image data that limits the performance of human or numerical observers on detection or estimation tasks \cite{park2005efficiency,rolland1992effect,barrett2013foundations}.
It is therefore important to have the capability of producing ensembles of NBPs that possess prescribed statistical properties associated with a specified to-be-imaged subject cohort; these NBPs can each be virtually imaged and, subsequently, ensemble-averaged objective IQ measures can be computed for use in assessing and refining USCT imaging technologies.
However, existing NBPs do not satisfy these requirements and are limited
by factors that include oversimplified anatomical structures\cite{wang2015waveform,huthwaite2011high,matthews2017regularized,li2009methodology,bakic2011development} or are representative of healthy subjects only \cite{lou2017generation, ali2019open}.  NBPs derived from clinical magnetic resonance images are available \cite{lou2017generation} but are severely limited in number; as such,  they do not accurately depict variability in breast anatomy or acoustic properties that will be present in a prescribed patient cohort.  
Other tools for generating NBPs\cite{li2009methodology,bakic2011development} rely on digital templates or segmented clinical images with simplified anatomical structures and consider only a limited number of tissue types. In summary, there remains an important need for developing NBPs for use in VITs of breast USCT that 1) comprise realistic structures and acoustic properties; 2) include lesions and/or other pathologies; and 3) are representative of the stochastic variability in breast size, shape, composition, anatomy, and tissue properties observed in a specified cohort of to-be-imaged subjects.
%
%

Recently, the Virtual Imaging Clinical Trials for Regulatory Evaluation (VICTRE) project~\cite{badano2018evaluation,badano2021silico} of the Food and Drug Administration (FDA) has validated and released software tools to generate realistic NBPs, as part of an end-to-end simulation framework for virtual mammography imaging studies. The breast size, shape, location, density, and extent of different tissues are tunable parameters, based on which stochastic and physically realistic three-dimensional (3D) numerical phantoms of tissue structures can be generated.  The tool also allows to embed a variety of lesions  (e.g.,  circumscribed or spiculated) at physiologically plausible locations.

In this work, a  methodology  for  producing  realistic  3D  numerical acoustic breast  phantoms  for  enabling  clinically-relevant VITs  of  USCT  breast  imaging  is  presented.
 This will be accomplished by extending the VICTRE NBPs for use in USCT, which will permit virtually imaging of ensembles of NBPs whose physical and statistical properties are representative of clinical cohorts.
Modifications to the VICTRE NBPs include: the determination of breast shape parameters consistent with a prone imaging position  \cite{ruiter2016analysis, duric2014clinical},
the stochastic assignment of tissue specific acoustic properties (density, SOS, and AA), as well as the modeling of acoustic heterogeneity within fatty and glandular tissues. 

To demonstrate the usefulness of the proposed computational framework, two case studies are presented. Case study 1 assesses the reconstructed SOS image quality using different compensation techniques to account for unknown AA. Case study 2 demonstrates the utility of the proposed framework for generating large-scale ensembles of NBPs for the training of deep learning-based USCT reconstruction methods. 
To accompany this work, a python library implementing the proposed approach for the generation of 3D acoustic phantoms has been made publicly available under GPL-2.0 \cite{Fu2021code}. Furthermore, two datasets have been publicly released under CC-0: The first consists of 52 2D breast phantom slices and corresponding USCT measurement data\cite{li2021NBPs2D}; The second contains 4 3D realizations of numerical breast phantoms\cite{li2021NBPs3D}.

The remainder of the paper is organized as follows. In Section \ref{sec:background}, background on USCT breast imaging and the FDA VICTRE project are provided. The stochastic generation of 3D anatomically and physiologically realistic numerical breast phantoms for USCT virtual imaging trials is introduced in Section \ref{sec:methods}. Several examples of NBPs generated with the proposed tool is presented in Section \ref{sec:examples}. Section \ref{sec:case_studies} contains the case studies that illustrate possible applications of the proposed phantoms to inform image reconstruction development. Finally, in Section \ref{sec:conclusions}, a discussion of the wide range of applications enabled by the proposed framework is provided.

\section{Background}
\label{sec:background}
\subsection{USCT breast imaging}

In recent decades, a number of research  groups  have  been  developing   USCT imaging technologies for breast imaging applications \cite{schreiman1984ultrasound,carson1981breast,andre1997high,duric2007detection}. In a typical breast USCT system, the patient lies prone on the imaging table and the breast to be imaged is submerged in water.  An array of ultrasound transducers surrounds the breast. Each transducer emits an acoustic pulse one by one until the breast is insonified from all directions. During each shot, all other transducers act as receivers, recording the transmitted, scattered, and reflected wavefield data.

Three types of USCT images are conventionally produced: reflectivity, SOS, and AA \cite{duric2018breast}.
Reflectivity images can be reconstructed by use of integral geometry-based approaches that are similar to the delay-and-sum methods widely employed in conventional B-mode imaging.
The majority of the SOS and AA reconstruction methods investigated to date are generally based on two categories: approximated wave equation methods \cite{simonetti2009diffraction,javaherian2020refraction,huthwaite2011high}, and full-waveform inversion (FWI) methods \cite{pratt2007sound,wiskin2020full,matthews2017regularized,wang2015waveform}.
Because FWI methods take high-order refraction and diffraction effects into account, they can produce images that possess higher spatial resolution images than those produced by use of linearized or approximate methods\cite{wiskin2012non,wang2015waveform,pratt2007sound}.
However, FWI is computationally expensive and memory burdensome, especially for 3D problems, thus hampering the widespread application of FWI to USCT breast imaging. Moreover, FWI suffers of the so-called \emph{cycle skipping phenomenon}\cite{witte2018full}, thus requiring an accurate initial estimate of the SOS map to ensure convergence to a useful solution.
As a result, there is still an imperative need to systematically investigate and optimize USCT reconstruction methods by means of computer-simulation studies.

\begin{figure}[tbh]
\centering
\includegraphics[width=1\linewidth]{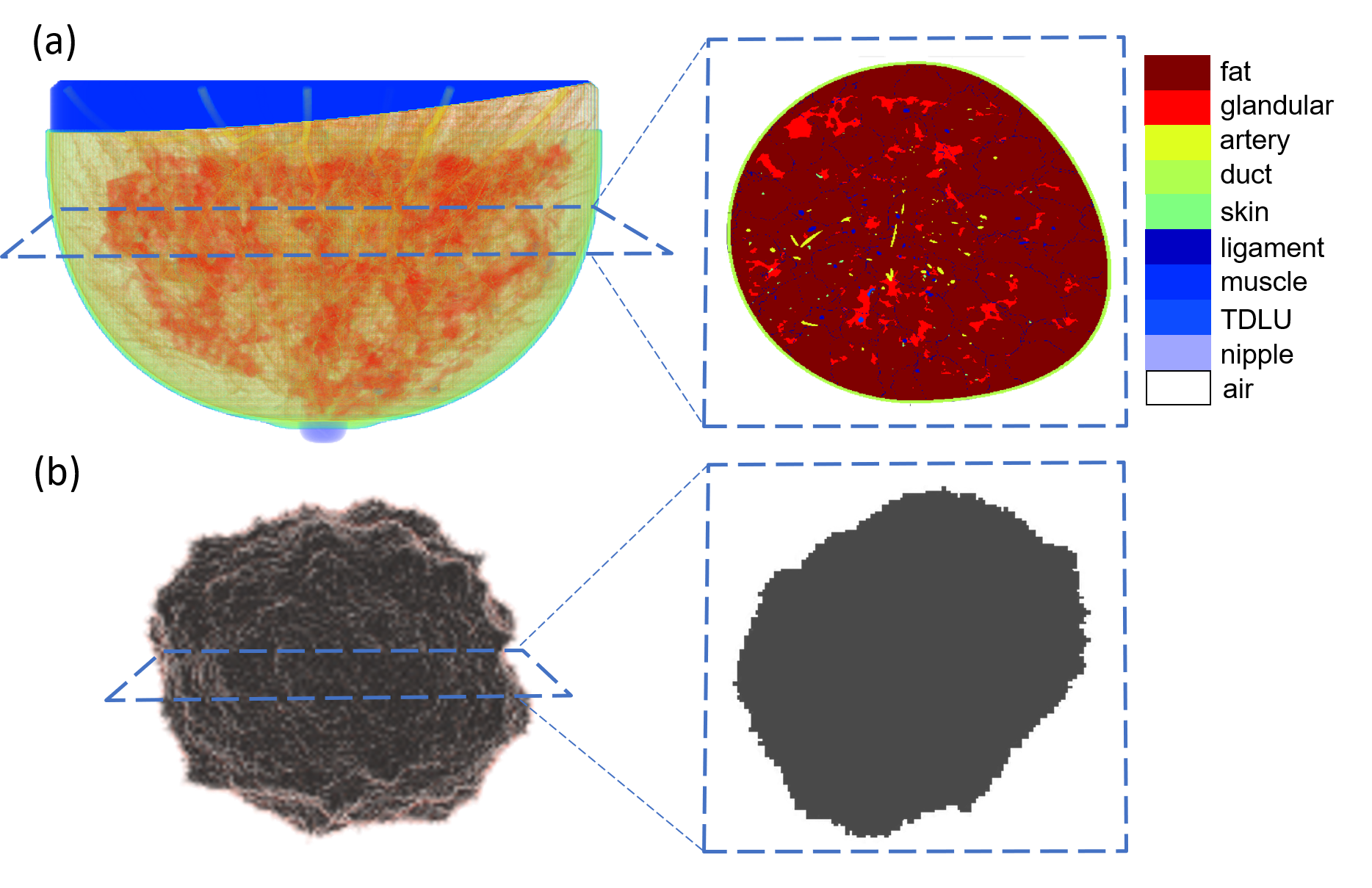}
\caption{ (a) Volume rendering of fatty breast phantom: partial transparencies are used to highlight anatomical structures, and cross-section view of this 3D breast phantom. (b) Volume rendering of spiculated lesion phantom. and cross-section view of this 3D lesion phantom.}\label{victre_ex}
\end{figure}%

\subsection{Description of VICTRE}
The Virtual Imaging Clinical Trials for Regulatory Evaluation (VICTRE) project of the Food and Drug Administration (FDA) has recently released a series of software tools to provide a complete simulated imaging chain for mammography and digital breast tomosynthesis  \cite{badano2018evaluation}.
The VICTRE software includes open source tools to generate three-dimensional (3D) random anthropomorphic voxelized phantoms of the human female breast \cite{graff2016new}. Using this tool, large ensembles of anthropomorphic numerical breast phantoms (NBPs) with realistic anatomical structures can be generated by specifying different virtual-patient characteristics that include breast type, shape, granularity, density, and size. By appropriate selection of physical attributes and material coefficients, the VICTRE NBPs can be customized for particular imaging tasks.


The VICTRE software generates NBPs corresponding to the four different levels of breast density defined according  to the American College of Radiology's (ACR) Breast Imaging Reporting and Data System (BI-RADS) \cite{american2013acr}: A) Breast is almost entirely fat, B) Breast has scattered areas of fibroglandular density, C) Breast is heterogeneously dense, and D) Breast is extremely dense. Each NBP is a 3D voxelized map consisting of ten tissue types: fat,  skin,  glandular,  nipple, ligament (connective tissue), muscle, terminal duct lobular unit, duct, artery, and vein. Large ensembles of stochastic NBPs with realistic variability in breast volume, shape, fraction of glandular tissue, ligament orientation, tissue  anatomy, can be generated by controlling input parameters and selecting the random seed number.

In addition, the VICTRE projects include tools to generate 3D numerical lesions phantoms (NLPs), which can be inserted into the NBPs at clinically plausible locations \cite{de2015computational}.  Two types of lesions, microcalcification clusters and spiculated masses, can be generated. The size and shape of the lesions can be customized. An example of anatomically realistic NBP and NLP generated using the VICTRE tools is shown in Fig. \ref{victre_ex}.

\begin{figure}[b]
\centering
\includegraphics[width=0.24\textwidth]{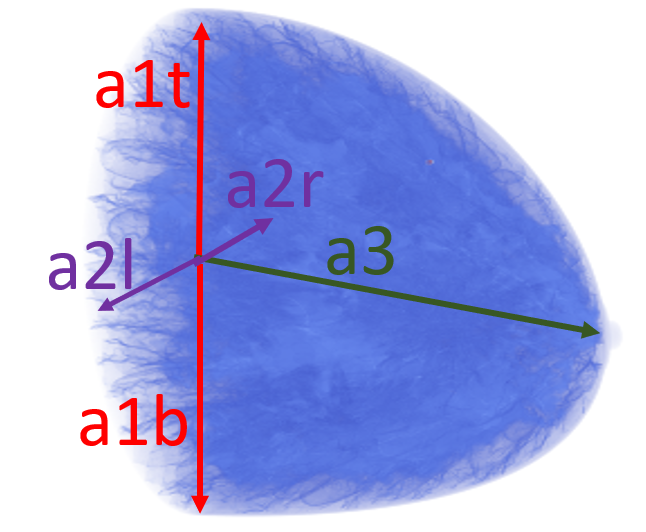}
\caption{Overview of size parameters: a1t, a1b, a2l, a2r, a3.}
\label{sizep}
\end{figure}
\begin{figure}
\centering
\includegraphics[width=0.4\textwidth]{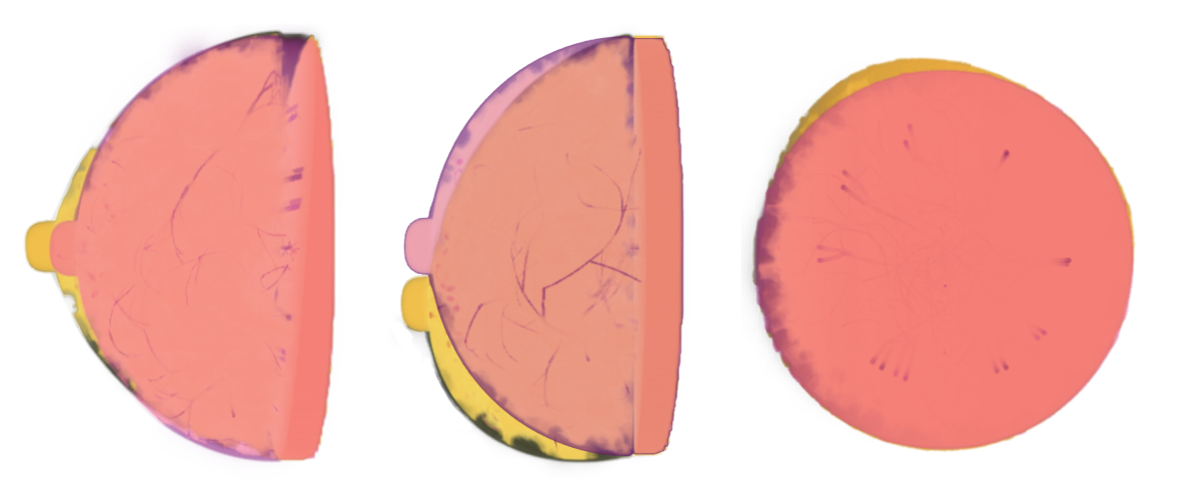}
\caption{Overview of deformation parameters. Red breast: hemispherical breast phantoms without deformation, yellow breast: deformed breast phantoms. \textbf{Left}: the effect of superquadric exponent deformation ($\epsilon_1$) in sagittal plane. \textbf{Center}: the effect of ptosis deformation ($B_0$, $B_1$) in sagittal plane. \textbf{Right}: the effect of turn top deformation ($H_0$, $H_1$) in coronal plane.}
\label{deformation}
\end{figure}

There exist several challenges that must be addressed
in order to extend the VICTRE project to produce NBPs for use in VITs of USCT technologies. These include determination  of  breast  shape parameters consistent with a prone imaging position, the stochastic assignment of tissue specific acoustic properties (density, SOS, and AA), and the modeling of acoustic heterogeneity within fatty and glandular tissues.

\section{Methods}
\label{sec:methods}
Several adaptations and customizations of the VICTRE tools were developed that
will enable the generation of large ensembles of  acoustic NBPs that display clinically relevant variability in both anatomical structures and acoustic properties.
The specific procedures for accomplishing this are described below.

\begin{table}[bth]
\centering
\caption{Shape and size parameters.} 
\tabcolsep=0.12cm
\begin{tabular}{|c|c|c| }
\hline
Parameters& Types A-C &Type D \\
\hline
$a_{1t}$ (cm)& $\mathcal{TN}$(5.85, 2.3275, 3.85, 7.70) & $\mathcal{TN}$(4.20,1.225,2.80,5.25)\\
 \hline
$a_{1b}/a_{1t}$ &  $\mathcal{N}$(1, 0.02) & $\mathcal{N}$(1, 0.02) \\
 \hline
$a_{2r}/a_{1t}$ &  $\mathcal{N}$(1, 0.05) & $\mathcal{N}$(1, 0.05) \\
 \hline
$a_{2l}/a_{2r}$  & $\mathcal{N}$(1, 0.05) & $\mathcal{N}$(1, 0.05) \\ 
 \hline
$a_{3}/a_{1t}$  & $\mathcal{TN}$(1.48, 0.18, 1, 1.6) &  $\mathcal{TN}$(1.22,0.1,0.75.1.5)\\
  \hline
$\epsilon_1$  & \multicolumn{2}{ c |}{$\mathcal{N}$(1, 0.1)}\\ 
   \hline
$B_0$  & \multicolumn{2}{ c |}{$\mathcal{TN}$(0, 0.1, -0.18, 0.18)}\\ 
   \hline
$B_1$  & \multicolumn{2}{ c |}{$\mathcal{TN}$(0, 0.1, -0.18, 0.18)}\\ 
   \hline
$H_0$  & \multicolumn{2}{ c |}{$\mathcal{TN}$(0, 0.15, -0.11, 0.11)}\\ 
   \hline
$H_1$ &  \multicolumn{2}{ c |}{$\mathcal{TN}$(0, 0.25, -0.3, 0.3)}\\ 
    \hline
 \end{tabular}
 \label{shape_dist}
\begin{flushleft}
$\mathcal{N}(\mu, \sigma)$: Gaussian distribution with mean $\mu$ and standard deviation $\sigma$.\\
$\mathcal{TN}(\mu, \sigma, a, b)$: Truncated Gaussian distribution in interval $(a,b)$.
\end{flushleft}
\end{table}

\subsection{Generation of anatomically realistic realizations of NBPs and lesion(s) insertion}
\label{sec:GenerationAnatomy}
The goal of this step is to generate large ensembles of anatomically realistic NBPs representing four different types of breast (extremely dense,  heterogeneously dense, scattered fibroglandular and fatty). Section \ref{sec:breast_shape_def} describes how shape and deformation parameters in the VICTRE NBPs can be set to generate virtual patients with variable breast sizes that are representative of a clinical population and shapes that are consistent with USCT imaging protocols. Section \ref{sec:relabeling} describes adaptations to the internal anatomical structures of the NBP to exclude tissues that are not relevant for USCT applications. Finally, Section \ref{sec:lesion_insertion} describes how one or more lesions are optionally inserted into the NBPs.

\subsubsection{Breast shape and deformation parameters}
\label{sec:breast_shape_def}
Appropriate distributions of breast size parameters were determined for each breast type based on clinical data \cite{huang2011characterization}.
In the VICTRE software, the shape of the breast is created by applying a series of transformations to a base superquadratic surface. A detailed description of the breast shape model was presented in \cite{chen2000modeling}. Here, the main parameters affecting size and shape of the breast are discussed. As shown in Fig. \ref{sizep}, the parameters $a_{1b}$, $a_{1t}$, $a_{2r}$, $a_{2l}$ adjust the breast volume in the top, bottom, left, right hemispheres, respectively. The parameter $a_3$ adjusts the length of the breast. Fig. \ref{deformation} illustrates how other parameters affect the final shape of the breast. The parameter $\epsilon_1$ is the quadric shape exponent along the polar angle. The ptosis deformation parameters $B_0$, $B_1$ model the sagging that affects a breast as a subject ages. Finally, the turn-pop deformation parameters $H_0$, $H_1$ change the shape of the top half of the breast laterally. This deformation allows the top part of the virtual breast to point towards the shoulder.
The probability distributions assigned to these parameters are summarized in Table \ref{shape_dist} and were set to be consistent with the patient lying prone on the examination table. Here---among all possible distribution with a specified mean, variance and having support in a bounded interval---a truncated Gaussian distribution ($\mathcal{TN}$) is chosen since it represent the maximum entropy distribution that satisfies such constraints.

\begin{figure}[tb]
\centering
\includegraphics[width=1\linewidth]{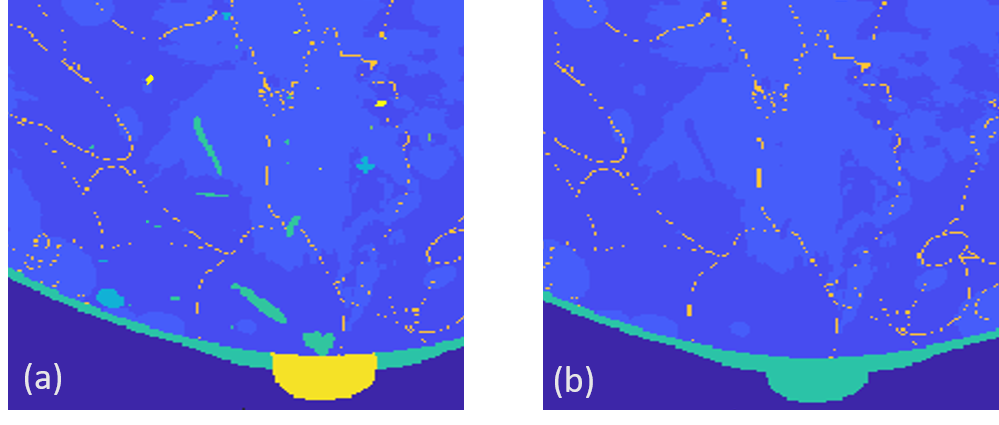}
\caption{Illustration of relabeling of tissues type invisible to USCT. (a): An anatomical phantom (tissue labels) generated by VICTRE. (b): A phantom after tissue relabeling. The different colors represent distinct tissue types.}  \label{label_rm}
\end{figure}%

\begin{table*}[bp]
\centering
 \caption{Acoustic property values of different tissue types.}
\begin{tabular}{| c | c | c | c | }
\hline
 Medium   &    SOS $[m/s]$  & AA $[Np/m/MHz^y]$ & Density $[kg/m^3]$ \\
 \hline
 Water          &    1500     @ $26^\circ$C       \cite{greenspan1959tables}                                                &0.025328436023 \cite{hasgall2018it} & 994 ~\cite{hasgall2018it}\\
 \hline
  Skin &  $\mathcal{TN}$(1555.0, 10.0, 1530, 1580) \cite{malik2016objective} & $\mathcal{N}$(21.158, 2.16)\cite{hasgall2018it}        &  $\mathcal{TN}$(1109, 14, 1100, 1125) \cite{hasgall2018it}\\
 \hline
 Fat            &   $\mathcal{TN}$(1440.2, 20.9, 1412, 1485)~\cite{malik2016objective,klock2016anatomy} & $\mathcal{N}$(4.3578, 0.436)\cite{hasgall2018it}                & $\mathcal{TN}$(911, 53, 812, 961) \cite{hasgall2018it}\\
 \hline
 Glandular  &  $\mathcal{TN}$(1540.0, 15.0, 1517, 1567 )~\cite{malik2016objective, klock2016anatomy}& $\mathcal{N}$(8.635, 0.86)\cite{hasgall2018it}   & $\mathcal{TN}$(1041, 45.3, 990, 1092)\cite{hasgall2018it}\\
 \hline
 Ligament   &  $\mathcal{TN}$(1457, 18.5, 1422, 1496)\cite{malik2016objective, klock2016anatomy}  & $\mathcal{N}$(14.506, 1.45)\cite{hasgall2018it}    &  $\mathcal{TN}$(1142, 45, 1110, 1174)\cite{hasgall2018it}\\
 \hline
Tumor     &  $\mathcal{TN}$(1548, 10.3, 1531, 1565) \cite{li2009vivo}& $\mathcal{N}$(31, 2.3)\cite{andre2013clinical}  & $\mathcal{TN}$(945, 20, 911, 999)\cite{sanchez2017estimating}\\
 \hline
  \end{tabular}
\label{tab1}
\end{table*}

\subsubsection{Relabeling of tissue types invisible to USCT}
\label{sec:relabeling}
The generated NBPs are high-resolution volumes with a voxel size as small as $50\,\mu{\rm m}$. Each voxel is assigned a label corresponding to one of the ten tissue types (fat,  skin, glandular, nipple, ligament, muscle, terminal duct lobular unit, duct, artery, and vein). Of these tissues, only four are typically visible in USCT imaging: fatty, glandular, skin, and ligament. Voxels corresponding to tissue types for which there is not enough clinical evidence that they can be well resolved in USCT imaging are relabeled as fatty or glandular based on the type of the neighboring voxels. An ad-hoc inpainting algorithm was designed to ensure consistent anatomical structures when relabeling voxels. The first step in the algorithm marks all voxels to be relabeled. Marked voxels are assigned to regions based on connectivity (two voxels are connected if they share a face) and process each connected region independently. For each region, the algorithm selects voxels near the boundary of the region (i.e. all voxels that share at least one face with unmarked voxels), reassigns their labels to the most occurring label among those of neighboring (unmarked) voxels, and unmarks them. This step is repeated until all voxels in all regions have been relabeled. 
An example of the result of replacement of USCT-invisible tissues is shown in Fig. \ref{label_rm}.

\subsubsection{Lesion insertion}
\label{sec:lesion_insertion}
To generate NBPs that contain tumors, synthetic lesions can be inserted in the healthy NBPs as follows. First, an ensemble of numerical tumor phantoms (NTPs) with various sizes and irregular (spiculated) shapes can be generated by the use of the VICTRE tool. One or more NTPs can then be inserted in each NBP at locations among those suggested by the VICTRE phantom tools as candidate tumor locations. Additional location constraints are included to ensure tumors do not overlap each other or skin layer and are not inserted too close to the chest or nipple.
\begin{figure}[tb]
\centering
\includegraphics[width=1\linewidth]{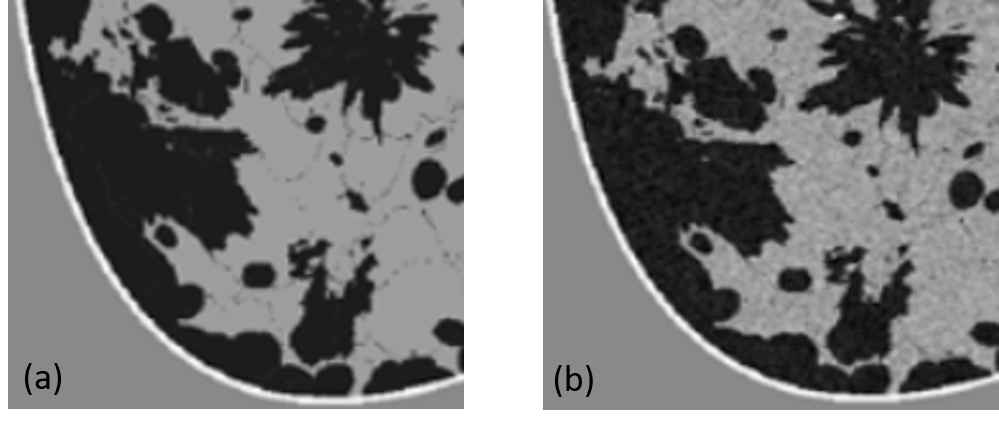}
\caption{ Illustration of texture generation on SOS phantoms. (a): A piecewise constant SOS phantom. (b): SOS phantom after texture generation.}\label{textuew}
\end{figure}%

\subsection{Assignment of acoustic properties}
\label{sec:acoustic_prop_maps}
By use of the anatomical breast  maps generated in Section \ref{sec:GenerationAnatomy},  3D acoustic NBPs can be established via stochastic assignment of acoustic properties.  The acoustic properties considered are the SOS $c$ ($m/s$), density $\rho(\mathbf{r})$ ($kg/m^3$), and AA coefficient $\alpha_0$ ($Np/m/MHz^y$) with power law exponent $y$. The 3D acoustic property maps are constructed as follows.
First,  acoustic properties values are stochastically assigned to each phantom voxel based on the tissue type label as described in Section \ref{sec:assigment_by_tissue_type}. Next, to model variations in the acoustic properties across voxels of the same tissue type, SOS and density maps are perturbed by additive coloured noise with a prescribed correlation structure as described in Section \ref{sec:sos_density_texture}. Finally, the choice of the power law exponent $y$ is presented in Section \ref{sec:att_exp}.

\subsubsection{Stochastic assignment of acoustic properties to each tissue type}
\label{sec:assigment_by_tissue_type}
Acoustic properties (SOS, AA, and density) are assigned to each voxel of the anatomical NBPs generated in Section \ref{sec:GenerationAnatomy} as follows. For each tissue type, values of SOS, AA, and density are sampled from a predefined probability distribution and assigned to all voxels of that tissue type. 
Table \ref{tab1} shows the probability distributions of the acoustic parameters assigned to each tissue type. These were chosen based on an comprehensive literature survey 
to represent anatomically realistic values. The 
SOS values of healthy breast tissues were based on the clinical studies reported in references \cite{klock2016anatomy,malik2016objective}. The distributions of density and AA in healthy breast tissues were set according to reference \cite{hasgall2018it}, a database providing comprehensive estimates of material properties of several human tissues, as well as statistical information about the spread  of those properties. This information was based on a meta-analysis of over 150 references. The variance of AA values for each tissue type was set to 10\% of the respective mean values. Finally, tumor acoustic properties were also chosen from clinical literature of  breast pathology \cite{andre2013clinical,sanchez2017estimating,li2009vivo}. 

Upon completion of this step, piecewise constant acoustic maps are constructed that present variability both in their values, which are randomly sampled, and spatial distribution, which is dictated by the NBP stochastic anatomical structure. 
Fig. \ref{textuew}(a) shows an example of a slice through a piecewise constant 3D SOS phantom generated by the described procedure.

\begin{figure*}[hb]
\centering
\includegraphics[width=0.9\linewidth]{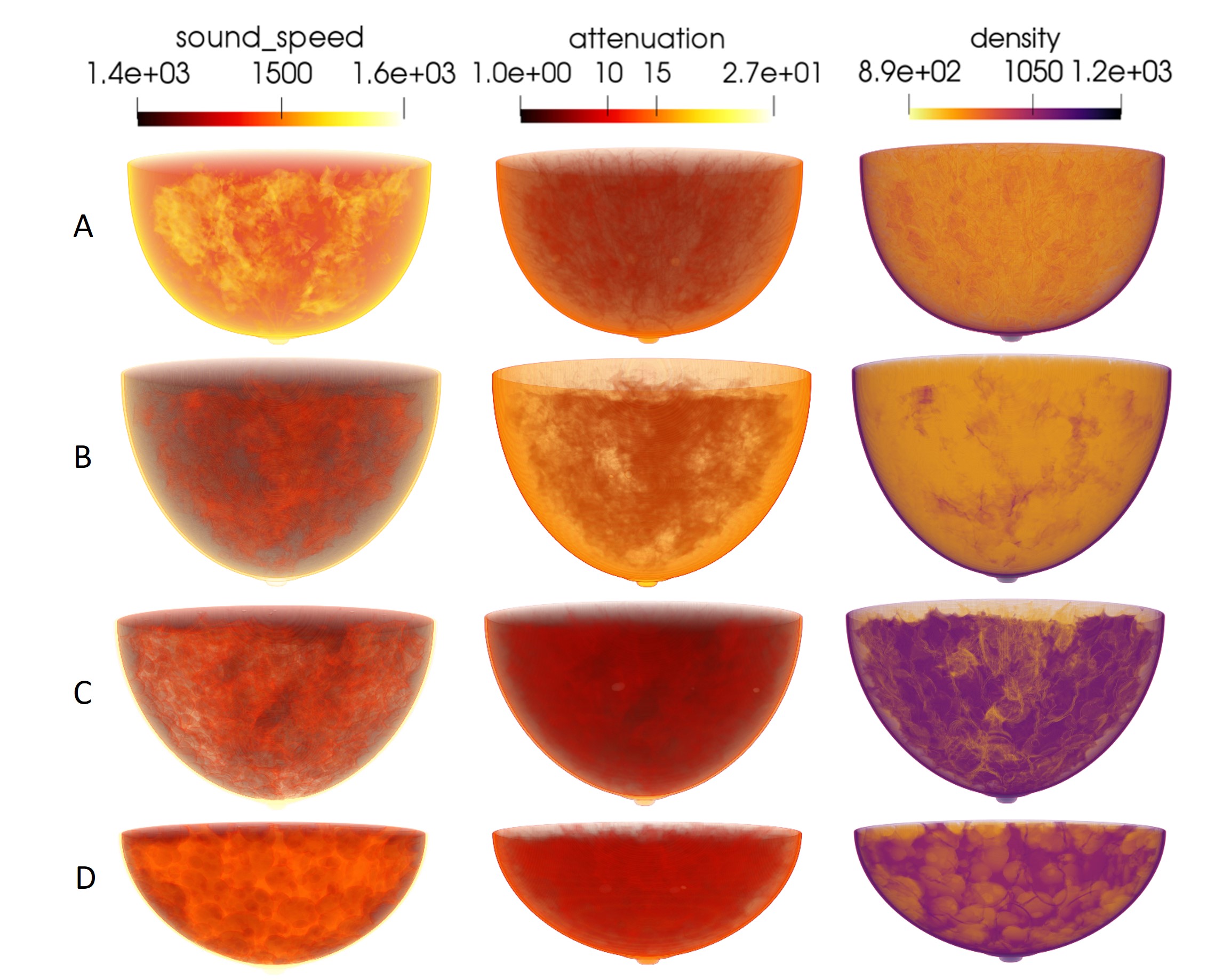}
\caption{3D rendering of acoustic phantoms from four breast types: From up to bottom:  (A) almost entirely fatty, (B) scattered areas of fibroglandular density, (C) heterogeneously dense, and (D) extremely dense. From left to right: the SOS $(m/s)$, AA $(Np/m/Mhz^y)$, and density $(kg/m^3)$ volumes.}\label{3dp}
\end{figure*}%
\begin{figure}[htb]
\centering
\includegraphics[width=0.92\linewidth]{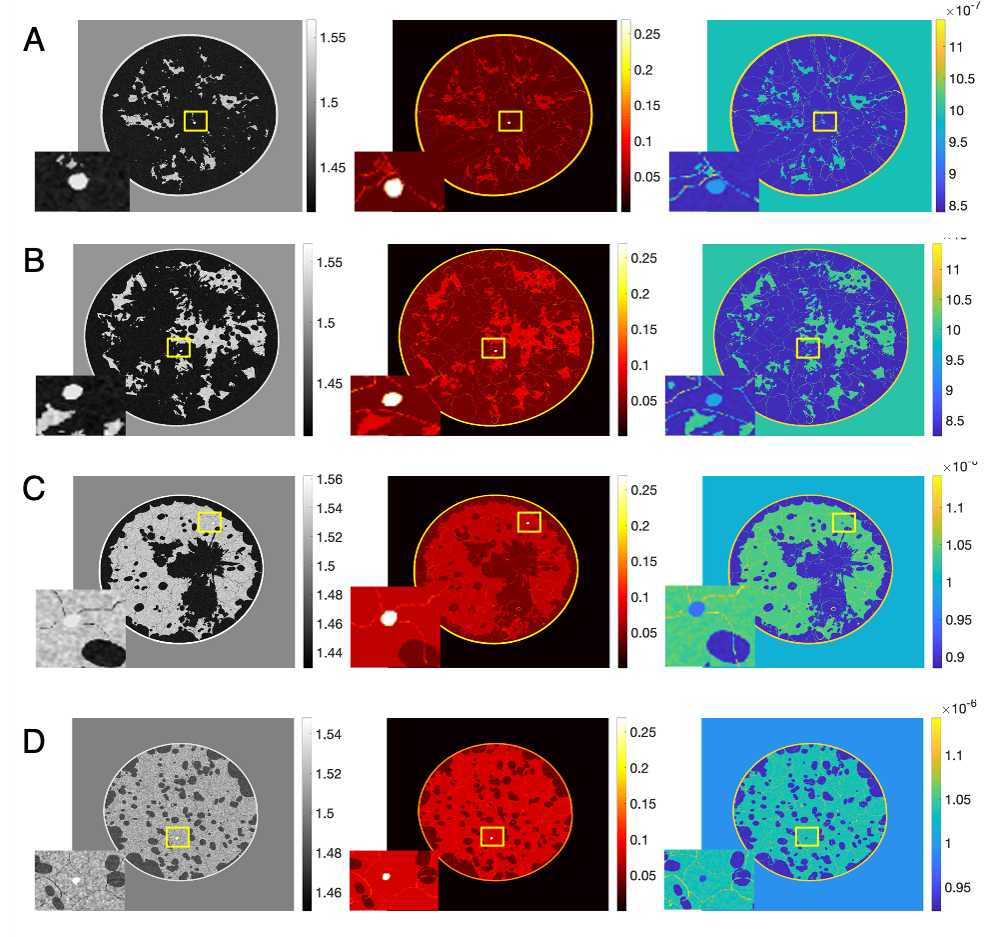}
\caption{A realization of a cross-sectional slice from four types breast.  From left to right: the SOS $(mm/\mu s)$ image, AA $(Np/m/Mhz^y)$ image and density $(kg/m^3)$ image. Tumor region is zoomed in. From up to bottom:  (A) almost entirely fatty, (B) scattered areas of fibroglandular density, (C) heterogeneously dense, and (D) extremely dense. }\label{dataset1}
\end{figure}%

\begin{table}[htb]
\centering
\caption{Pointwise standard deviations and correlation lengths uses to model texture in fatty and glandular tissues.}
\label{tab:texture_properties}
\begin{tabular}{|c|c|c|c|c|}
\hline
\multirow{2}{*}{Property} & \multicolumn{2}{c|}{SOS} & \multicolumn{2}{c|}{Density} \\ \cline{2-5} 
           & $\sigma\, (m/s)$& $\ell\, (mm)$ & $\sigma\, (kg/m^3)$ & $\ell\, (mm)$ \\ \hline
Fatty tissue$^*$  &  28.8        & 0.21       &  18.22 & 0.21       \\
Glandular tissue &  30.4         & 0.21       & 20.82   & 0.21       \\ \hline
\end{tabular}\\[1mm]
$^*$ Random texture in fatty tissues is truncated within the $\pm 0.9\sigma$ range.
\end{table}

\subsubsection{Modeling spatial heterogeneity within fatty and glandular tissues}
\label{sec:sos_density_texture}
Acoustic scattering in breast tissues arises not only from jumps in acoustic impedance across tissue types, but also from spatial heterogeneity within each tissue \cite{insana1990describing}. The latter is a predominant effect in fatty and glandular tissues. To account for the spatial heterogeneity within these tissues, random textures are introduced into the SOS and density maps.  SOS and density textures in glandular tissue are modeled as a spatially correlated Gaussian random field with zero mean and Gaussian covariance function. SOS and density textures in fatty tissue are modeled as truncated (plus or minus 0.9 standard deviations) spatially correlated Gaussian random field with zero mean and Gaussian covariance function account for the lower acoustic scattering observed in fatty tissues\cite{franceschini20062}.   The pointwise standard deviations $\sigma$ and correlation lengths $\ell$ are shown in Table \ref{tab:texture_properties}  and are based on reflectivity tomography studies~\cite{franceschini20062}. SOS and density textures are sampled independently one from the other. Each voxel in the generated textures maps is added to the corresponding voxel in the piecewise constant property maps described in the preceding paragraph; this results in NBPs that display random heterogeneity with the glandular and fatty tissues.
Fig. \ref{textuew}(b) shows an example of a slice through a 3D SOS phantom that contains tissue texture generated by the described procedure.
 Note how acoustic heterogeneity is stronger in glandular tissue (gray regions) than in fatty tissue (black regions).

\subsubsection{Power law attenuation model}\label{sec:att_exp}
To model frequency dependence in AA, a fractional power law model \cite{szabo1995causal} is assumed. Specifically, frequency dependent AA $\alpha$ ($Np/m$) is defined as
\begin{equation}
   \alpha = \alpha_0 f^y,
\end{equation}
where $\alpha_0$ ($Np/m/MHz^{-1}$) is the AA coefficient, $y$ is the fractional power law exponent, and $f$ ($MHz$) is the acoustic wave frequency. In general, the exponent $y$ varies for different tissue types and estimates for several breast tissues can be found in the IT’IS database~\cite{hasgall2018it}. However, several widely employed time-domain wave propagation solvers\cite{treeby2012modeling,komatitsch2018specfem} assume a spatially homogeneous exponent $y$.

To address this, a homogenization technique based on the solution of a nonlinear least squares problem is proposed.  The proposed technique considers wave propagation in one spatial dimension for which an analytical model of AA can be constructed. Specifically, for a monochromatic wave  with frequency $f$ propagating through a heterogeneous medium with thickness $L$, the log amplitude ratio $\ell(f)$ between the transmitted $A_t$ and incident $A_i$ wave is  
\begin{equation}
\frac{A_t}{A_i} = e^{ -\ell(f) }\quad {\rm with} \quad \mu(f)=\int_L \alpha_0(x) f^{\tilde{y}(x)}dx, 
\end{equation}
where $\tilde{y}(x)$ is the tissue-dependent fractional power law exponent.
Since attenuation in water is negligible and the volume  of skin, tumor, and ligament tissues is small compared to the whole breast, a medium consisting of only fatty and glandular tissues is considered. Under this simplifying assumption, $\ell(f)$ is determined as a function of the fatty tissue volume fraction only.
The spatially homogeneous fractional power law exponent $y$ is then defined as
\begin{equation}
y = \operatornamewithlimits{argmin}_y \sum_k \left(\ell(f_k) -  \overline{\alpha_0} f_k^{y}\right)^2,
\label{eq:homo_exp}
\end{equation}
where $\overline{\alpha_0}$ is the average value of $\alpha_0(x)$ and the frequencies $f_k$ ($k=1,\ldots, K$) are uniformly distributed over the range of frequencies typically employed in USCT imaging. 
Table \ref{y_estimate_breast_type} reports the estimated power law exponent $y$ as a function of the fatty tissue volume fraction $v_{\rm fat}$, when $K= 22$ frequencies evenly spaced between 0.2 MHz and 2.3 MHz  are used to evaluate Eq.\ \eqref{eq:homo_exp}.

\begin{table}[htp!]
\centering
\caption{Homogeneous power law exponent $y$ as a function  of the fatty tissue volume fraction $v_{\rm fat}$.}
\begin{tabular}{| c | cccc |}
\hline
Breast type & A&B&C&D \\
\hline
$v_{\rm fat}$ & $\sim 95\%$ & $\sim 85\%$ & $\sim 66\%$&  $\sim 40$\% \\
\hline
 $y$ &1.1151 &1.1642 &1.2563 & 1.3635 \\
  \hline
  \end{tabular}
  \label{y_estimate_breast_type}
\end{table}

\begin{figure}[tb]
\centering
\includegraphics[width=0.49\textwidth]{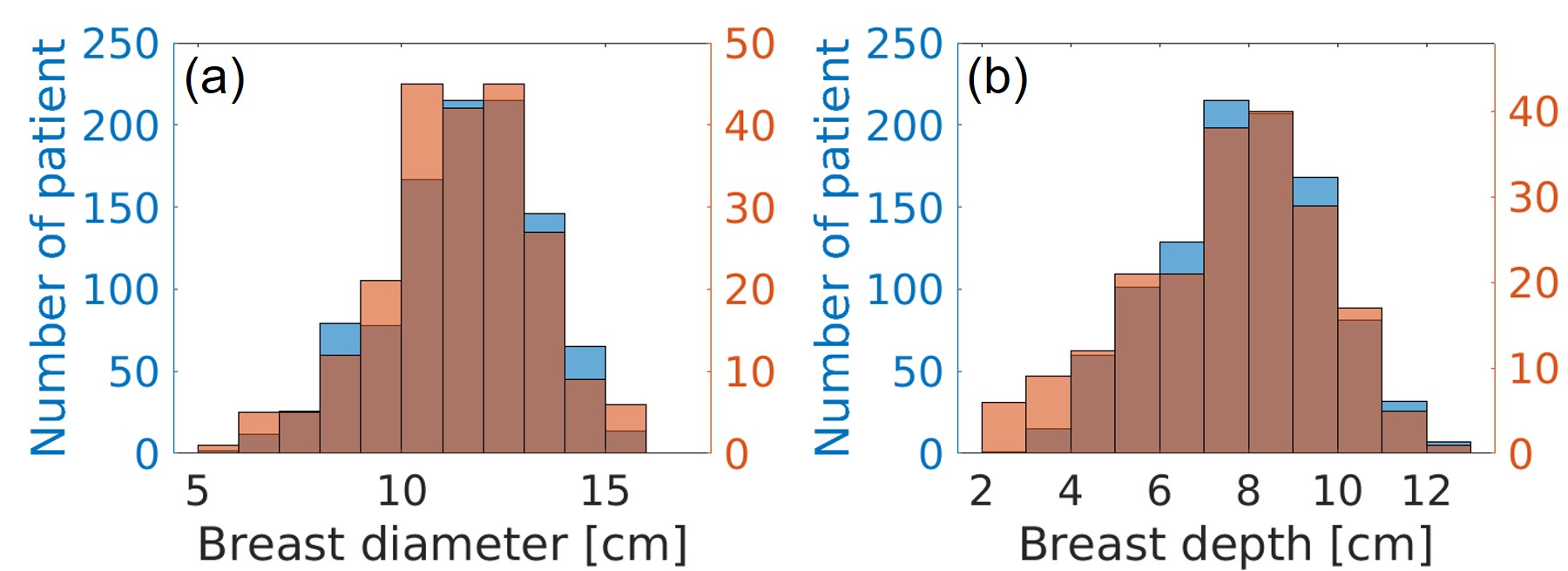}
\caption{Breast size distribution comparison. a): The breast diameter distribution. b): The depth  distributions. Blue: distributions of the generated NBPs. Orange: distributions estimated from clinical data~\cite{huang2011characterization}.}
\label{sizedist}
\end{figure}

\section{Examples of generated NBPs}
\label{sec:examples}

Fig. \ref{3dp} shows four 3D visualization examples, one for each breast type, of 3D acoustic NBPs produced by the proposed framework. The generation of the anatomical structures for these NBPs using the VICTRE tools took about 80$\sim$240 minutes on a single node of the Golub cluster at the University of Illinois at Urbana-Champaign campus cluster (Two 10-core Intel E5-2670v2 CPUs and 64 GB of memory per node). Tissue relabeling, tumor insertion, and assignment of spatially varying acoustic properties took between 20 and 50 minutes on the same machine. Time variations depended on the volume of phantoms. Paraview\cite{ahrens2005paraview} was used for volume rendering to highlight internal structures. Note the variability in size, shape, internal structures, and values of acoustic properties among the four NBPs. Fig. \ref{dataset1} shows examples of 2D cross-sectional slices extracted from the phantoms, one for each breast type. Yellow rectangles indicate the location of the inset zoom region where a lesion was inserted. The diameters of the inserted lesions were sampled from a uniform distribution between 1.5 mm and 5 mm, to mimic small lesions in early breast cancer. 

Fig. \ref{sizedist} compares the distributions of the breast diameter and depth in a virtual population of 1,000 NBPs to that observed in a sample of 219 women with age ranging between 35 and 82 years and median age of 54 years \cite{huang2011characterization}. The proportion for each breast type in the virtual population was set to 10\% for breast types A and D, and 40\% for breast types B and C\cite{american2013acr}. The figure shows good qualitative agreement in the diameter and depth distributions between the virtual population and the clinical sample. It is worth noting that the distributions of the virtual population are skewed towards slightly larger breast sizes compared to those of the clinical sample. This is intentional and aims to address a limitation of the sample in \cite{huang2011characterization}, which is biased towards denser---and therefore smaller---breast types (23\% type A, 40\% type B, 28\% type C, and 9\% type D).

\section{Case studies}
\label{sec:case_studies}
Two case studies  were conducted to demonstrate the usefulness of the proposed framework for generating acoustic NBPs. 
Case study 1 (Section \ref{sec:cs_cs1}) assesses  reconstructed SOS image quality when heuristic procedures for compensating for unknown AA are employed. Case study 2 (Section \ref{sec:cs_cs2}) demonstrates the utility of the proposed framework  for  the  training and assessment of  deep  learning-based  USCT  reconstruction methods. In both studies, 2D cross-sectional slices extracted from the 3D NBPs are (virtually) imaged using the stylized 2D imaging system described in Section \ref{sec:2DvirtualImagingSystem}. 


\subsection{Virtual imaging system}
\label{sec:2DvirtualImagingSystem}
A stylized 2D virtual imaging system was modeled to generate USCT measurement data. It comprised 1024  idealized, point-like, transducers that were evenly arranged in a circular array with a radius of 110 mm.  The excitation pulse employed in this study was assumed to be spatially localized at the emitter location. The central frequency and duration of the pulse were set to 1 MHz and 10 $\mu$s, respectively. The pulse profile $s(t)$ was defined as the sum of three sinusoidal functions tapered by a Gaussian kernel as
\begin{equation}
\begin{split}
s(t) = &\exp\left(-\frac{(t-t_s)^2}{2\sigma^2}\right)\\
{}&\times\left(\frac{1}{8}\sin\pi f_c t+\sin 2\pi f_c t+ \frac{1}{8}\sin 4\pi f_c t\right),
\label{masource}
\end{split}
\end{equation}
where $\sigma=1.6 \, \mu s$ is the standard deviation of the Gaussian kernel, $t_s = 3.2 \, \mu s$ is a constant time shift, and $f_c = 1$ MHz is the central frequency. The maximum frequency of $s(t)$ is $2.3$ MHz.

Cross-sectional slices were extracted from the 3D NBPs and centered within the field of view of the  imaging system. 
Bilinear interpolation was employed to downsample the maps of acoustic properties to a computational grid comprised of 0.1 mm isotropic pixels. 
To emulate the imaging process, the propagation of the pressure waves through the object was modeled by solving the lossy acoustic wave equation with power law frequency-dependent AA \cite{treeby2010modeling} by use of a time-explicit pseudo-spectral k-space method \cite{tabei2002k,huang2013full,matthews2018parameterized}. Further details regarding the wave solver and its implementation are presented in Appendix \ref{app:k-space-solver}. 
The simulated measurement data were corrupted with Gaussian i.i.d. noise that had zero mean and a standard deviation of 0.02\% of the maximum pressure amplitude at the emitting transducer. 

Computation of USCT measurement data for a single slice took about 110 GPU hours using a single NVIDIA GK110 Kepler GPU on the Blue Water cluster at the National Center for Super computing Applications.

\begin{table}[htp!]
\caption{Reference SOS reconstructions: mean square error (MSE) and structural similarity index measure (SSIM)}
\label{tab:ref_recon}
\centering
\begin{tabular}{|c|c|c|}
\hline
Breast type & MSE (std) & SSIM (std) \\
\hline
A & 1.786e-04 (3.923e-5) & 0.9835 (0.0056)  \\
B & 2.571e-04 (1.087e-4) & 0.9788  (0.0069) \\
C & 2.459e-04 (1.797e-4) & 0.9732 (0.0102) \\
D  & 2.258e-04 (1.301e-4) & 0.9835 (0.0066) \\
\hline
all types &2.269e-04 (1.210e-04) &  0.9799 (0.0083)\\
\hline
\end{tabular}
\end{table}

\begin{table*}[htb]
\centering
\caption{Case study 1: MSE and SSIM of SOS images using the TRAM and DDAC approaches to compensate for unknown AA.}
\label{tab:recon}
\begin{tabular}{|c|c|c|c|c|c|c|}
\hline
\multicolumn{2}{|l|}{} &
  \multicolumn{1}{c|}{A} &
  \multicolumn{1}{c|}{B} &
  \multicolumn{1}{c|}{C} &
  \multicolumn{1}{c|}{D} & All types \\ \hline
\multirow{2}{*}{MSE (std)} 
& TRAM & \textbf{1.825e-4} (3.863e-5)  & \textbf{2.853e-4} (1.199e-4)&2.686e-4 (1.449e-4) & \textbf{2.300e-4} (2.006e-4)  & 2.433e-4 (1.414e-4) \\ \cline{2-7} 
& DDAC & 3.132e-04 (7.309e-5) & 3.521e-4 (1.210e-4) & 2.668e-4 (1.666e-4) & 2.720e-4 (2.059e-4) & 3.006e-4 (1.513e-4) \\ \hline
\multirow{2}{*}{SSIM (std)} 
& TRAM & \textbf{0.9819} (0.0080) & \textbf{0.9766} (0.0083) & 0.9700 (0.0126) &  \textbf{0.9843} (0.0066) & 0.9785 (0.0103)\\ \cline{2-7} 
& DDAC & 0.9777 (0.0058) &0.9732 (0.0073) & 0.9709 (0.0094)  & 0.9818 (0.0071) &  0.9761 (0.0085)\\ \hline
\end{tabular}\\
\end{table*}

\subsection{SOS images reconstructed under favorable conditions}\label{sec:cs_references}
Reconstruction of SOS images under favorable conditions (namely, AA map known exactly) are considered here. However, this study does not represent an inverse crime as it includes three sources of model mismatch: 1) A constant density map was employed in the reconstruction; 2) Measurement data were corrupted with additive Gaussian noise; 3) Reconstruction were performed on a coarser grid.
By use of the procedures described above, 2D slices from 52 NBPs (13 for each of the 4 breast types) were extracted and virtually imaged to produce USCT measurement data. From these data, SOS images were reconstructed on a grid with pixel size of 0.2 mm by use of a previously published waveform inversion with source encoding (WISE) method\cite{wang2015waveform}. The reconstruction was initialized by use of a blurred version (Gaussian blur with 8 mm correlation length) of the true SOS. 
Because true values of AA were considered here, the reconstructed SOS estimates are expected to generally be of higher quality than would be obtained if attenuation properties had to be concurrently estimated with the SOS, or if incorrect fixed values of AA were employed.
In this sense, it will be useful to compare these reconstructed SOS estimates 
against the images reconstructed in the two case studies below.

\begin{figure}[htp!]
\centering
\includegraphics[width=1\linewidth]{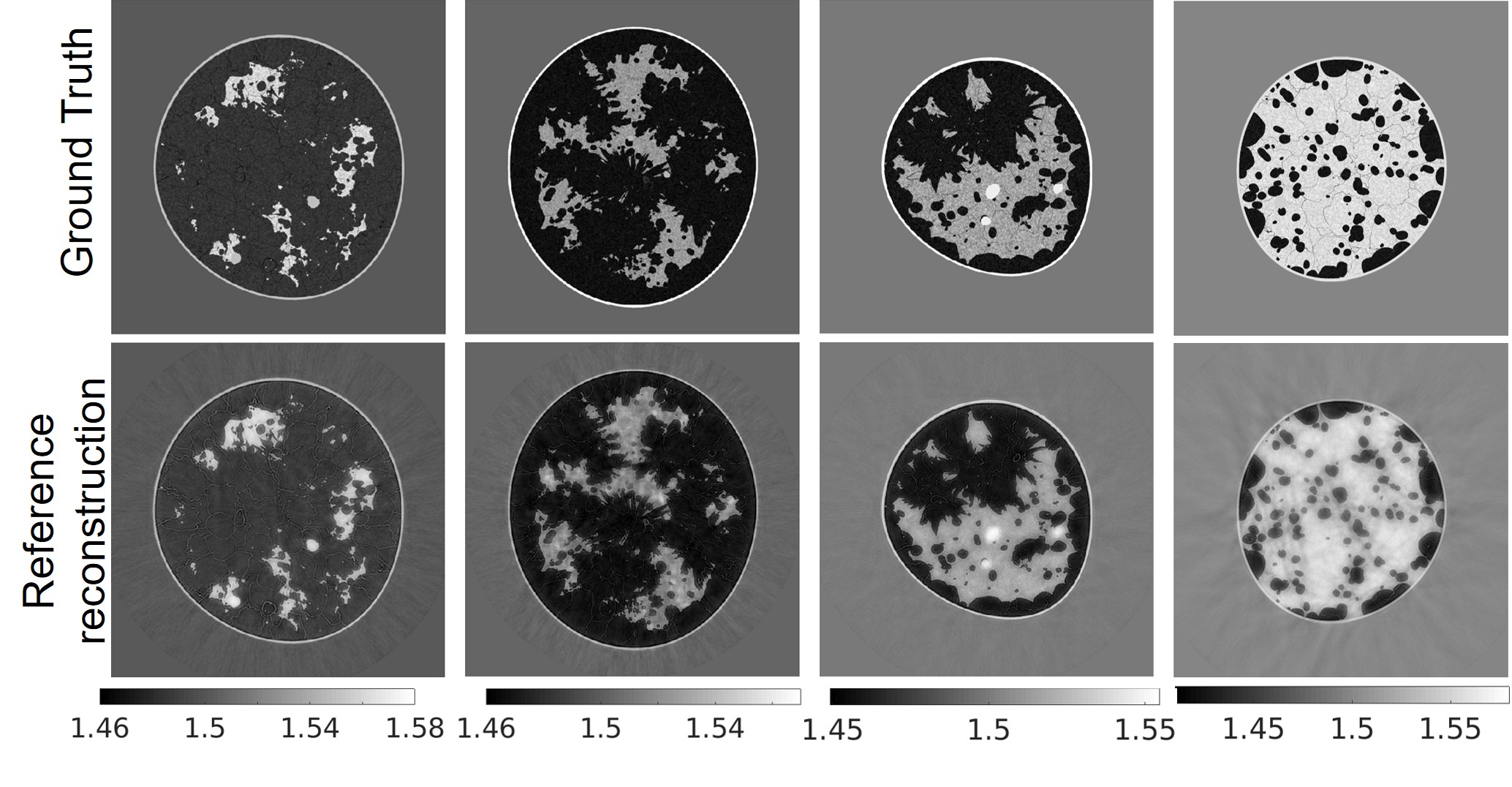}
\caption{Reference images: Ground truth (top row)  and reconstructed (bottom row) SOS maps. From left to right: breast type A-D. The unit is ($mm/\mu s)$.}
\label{fig:ref_recon_examples}
\end{figure}

Fig. \ref{fig:ref_recon_examples} presents examples, one for each breast type, of the ground truth and  reconstructed SOS images assuming the AA distribution and density are known. Table \ref{tab:ref_recon} reports the average mean square error (MSE)\footnote{Only pixels within the breast region were used to evaluate the MSE.} and structural similarity index measure (SSIM) \cite{wang2004image} for each breast type.

\subsection{Case study 1: Heuristic compensation of AA in SOS reconstruction}\label{sec:cs_cs1}
In this study, two heuristic approaches to compensating for AA when reconstruction SOS estimates were compared: 
a two-region attenuation model (TRAM) and a data domain attenuation compensation (DDAC). 
\begin{figure}[htp!]
\centering
\includegraphics[width=1\linewidth]{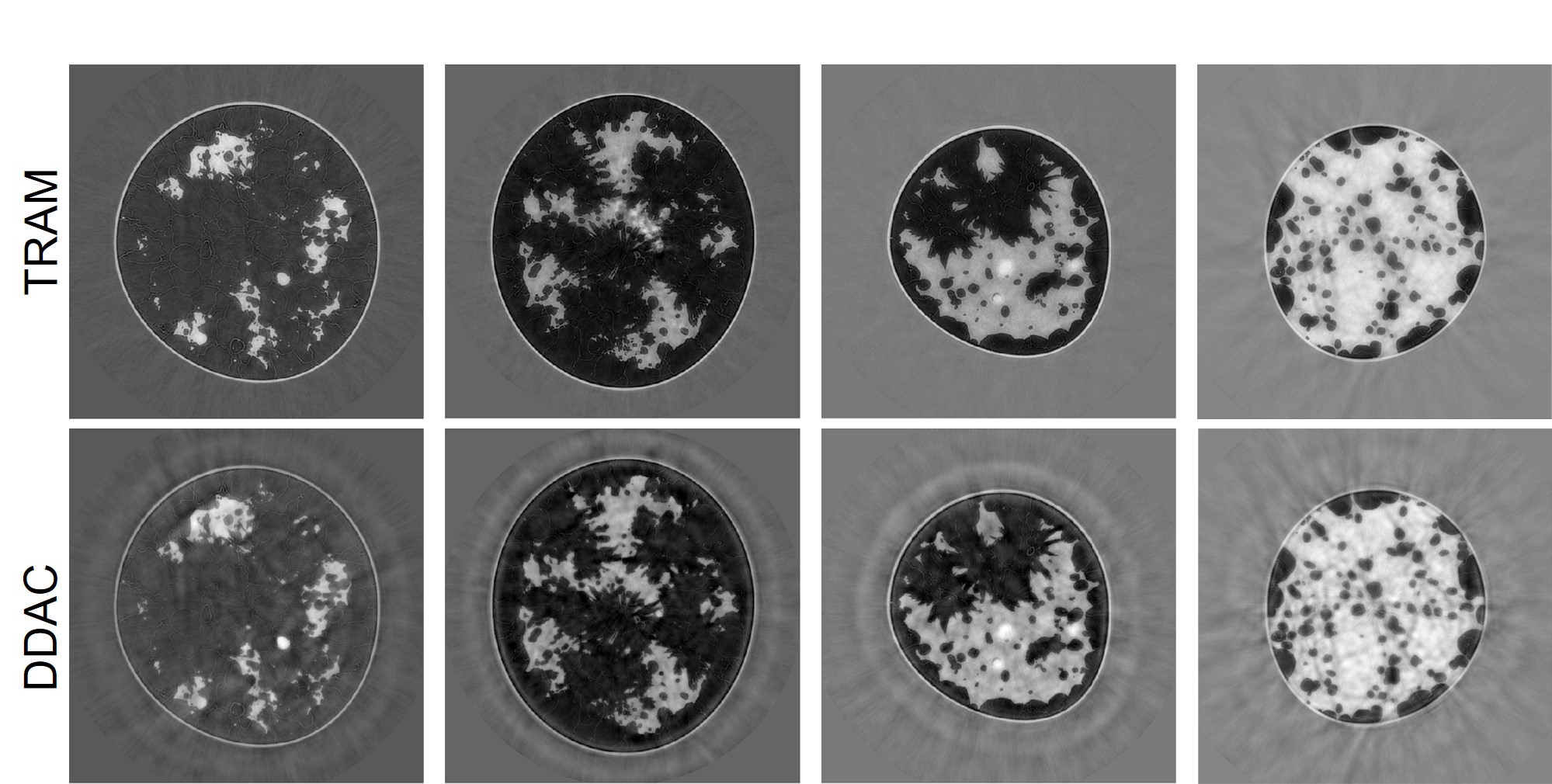}
\caption{Case study 1: Reconstructed SOS images corresponding to the same phantoms shown in Fig. \ref{fig:ref_recon_examples} using TRAM (top row) and DDAC (bottom row). From left to right: breast type A-D. The unit is ($mm/\mu s)$.} \label{fig:cs1_recon}
\end{figure}
The TRAM assumes that the breast boundary is known (reflectivity imaging could be possibly used to estimate it) and assigns one constant AA value to the water bath ($\alpha_0=0$) and another to the breast region. The attenuation coefficient of the breast region was set to 5.20 $[Np/m/MHz^y]$, which corresponds to a weighted average (80\%-20\% split) of the mean values of AA in fatty and glandular tissues as reported in Table \ref{tab1}. 
The heuristic DDAC procedure seeks to compensate for AA by modifying the amplitudes of the recorded pressure data, rather  than  explicitly modeling  attenuation in the wave propagation forward model. Specifically, for each pair of emitting/receiving transducers, the maximum amplitude of the recorded  signal was re-scaled to match that of the corresponding measurement when only the water bath was present\cite{matthews2017image}.
The generation of synthetic data and reconstruction method used in this case study are the same as described in Section \ref{sec:cs_references}.

\begin{figure}[htp!]
\centering
\includegraphics[width=0.99\linewidth]{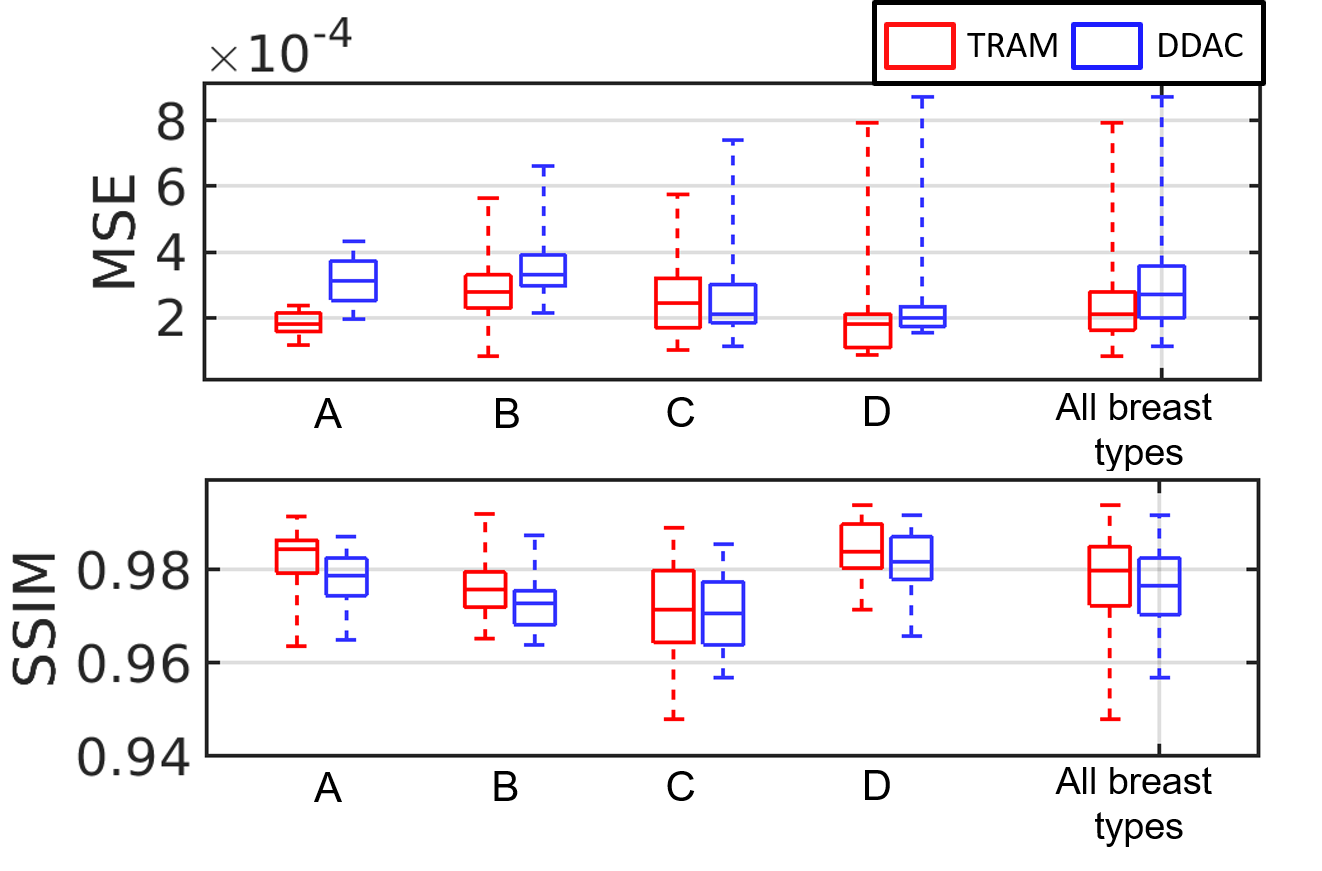}
\caption{Case study 1: Boxplots of MSE and SSIM value with respect to TRAM and DDAC. From left to right: breast types A-D and all breast types together.}
\label{fig:box_recon}
\end{figure}%

Fig. \ref{fig:cs1_recon} shows examples of reconstructed images of four breast types using the proposed techniques (TRAM and DDAC) to compensate for unknown AA properties. The corresponding ground truth images and reference reconstructions are shown in Fig. \ref{fig:ref_recon_examples}. Table \ref{tab:recon} shows quantitative evaluations of all reconstructed images on each breast type from A-D. Fig. \ref{fig:box_recon} shows the variation with respect to MSE and SSIM in all image samples and breast types A-D.
In all the cases tested but 9, TRAM led to measurable improvement (i.e. smaller MSE) compared to DDAC. Remarkably, of the 9 cases in which DDAC led to a smaller MSE, 6 were for breast type C. Furthermore, TRAM achieved a MSE smaller than that of the reference reconstruction (Section \ref{sec:cs_references}) for 25 out of the 52 tested cases, thus suggesting that TRAM is effective approach to compensate for unknown AA in FWI reconstruction of SOS maps.

\vspace{-0.1cm}
\begin{figure}[tb]
\centering
\bfseries\includegraphics[width=1\linewidth]{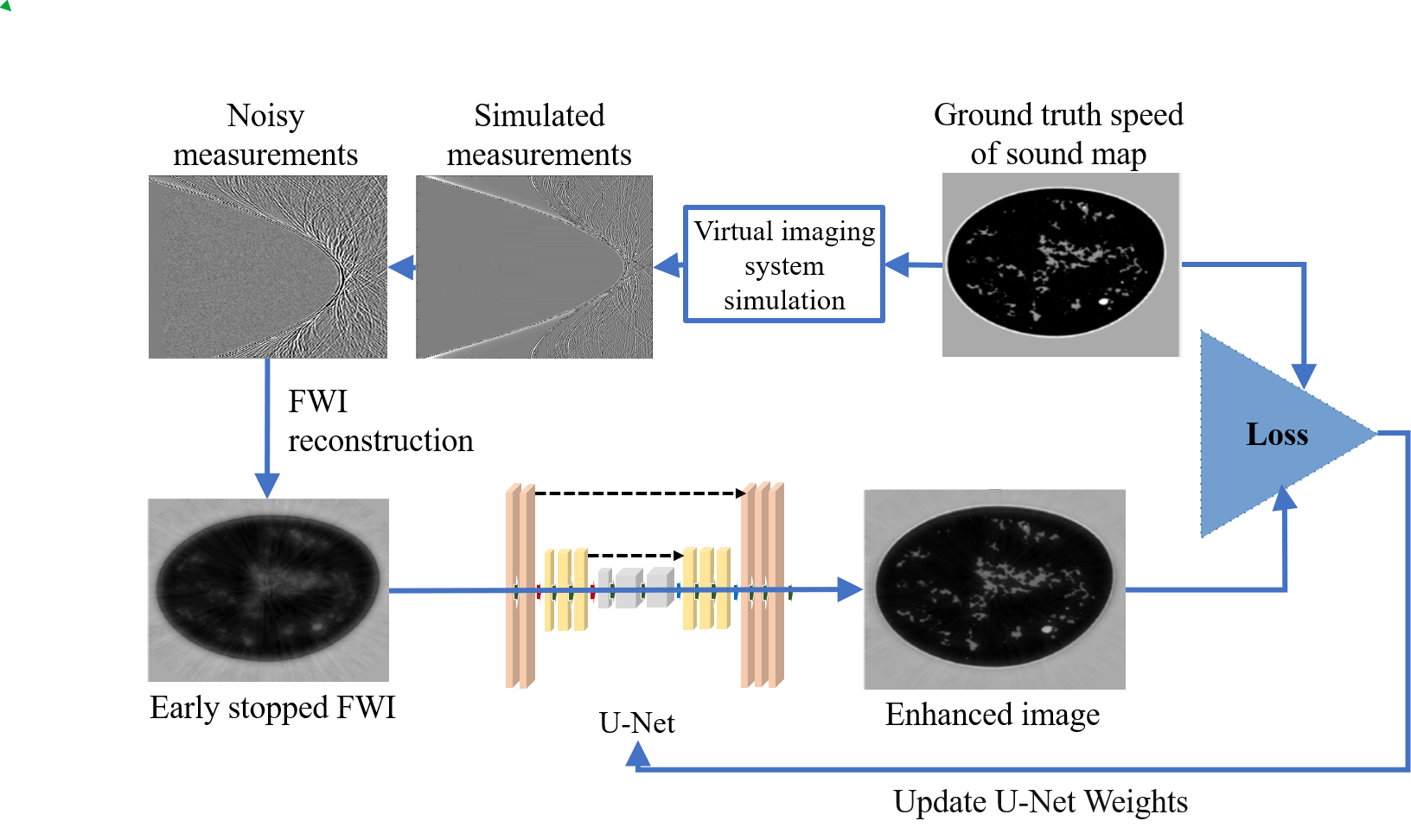}
\caption{Case study 2: The supervised deep learning framework for SOS reconstruction.}\label{dlframe}
\end{figure}%

\subsection{Case study 2: Deep learning reconstruction method}\label{sec:cs_cs2}
There remains an important need to lessen the computational burden of FWI. A supervised learning-based method is proposed to reduce the number of FWI iterations and drastically lessen the computational burden, as well as enhancing the reconstruction by using ground truth images as training data. Fig. \ref{dlframe} illustrates the proposed learning framework, in which a deep neural network (a 5-level U-Net \cite{ronneberger2015u}) is trained to minimize the mean square error of the reconstructed SOS images. The input to the network is an intermediate SOS estimate  obtained by early stopping of the WISE method after 35 iterations. The rationale of this method is that early-stopped reconstructed images capture structural information of the SOS map but lack quantitative accuracy. The network was trained for 220 epochs on a dataset consisting of 622 2D slices, each extracted from a different NBP (312 type B NBPs and 310 type C NBPs). The Adam optimizer was used with a batch size equal to 32 and the initial learning rate 0.001. The learning rate was reduced by a factor of 0.9 after each epoch. Several models corresponding to different architecture hyper-parameters (e.g., number of layers at each level) were trained. The selected model was that achieving the highest mean MSE on the validation set consisting of 100 slices (equally split between types B and C). 

The testing set consisted of the 52 2D phantoms described in Section \ref{sec:cs_references}, thus allowing us to evaluate the accuracy of the network for both in-distribution (types B, C) and out-of-distribution data (types A, D).

\begin{figure}[htb]
\centering
\includegraphics[width=1\linewidth]{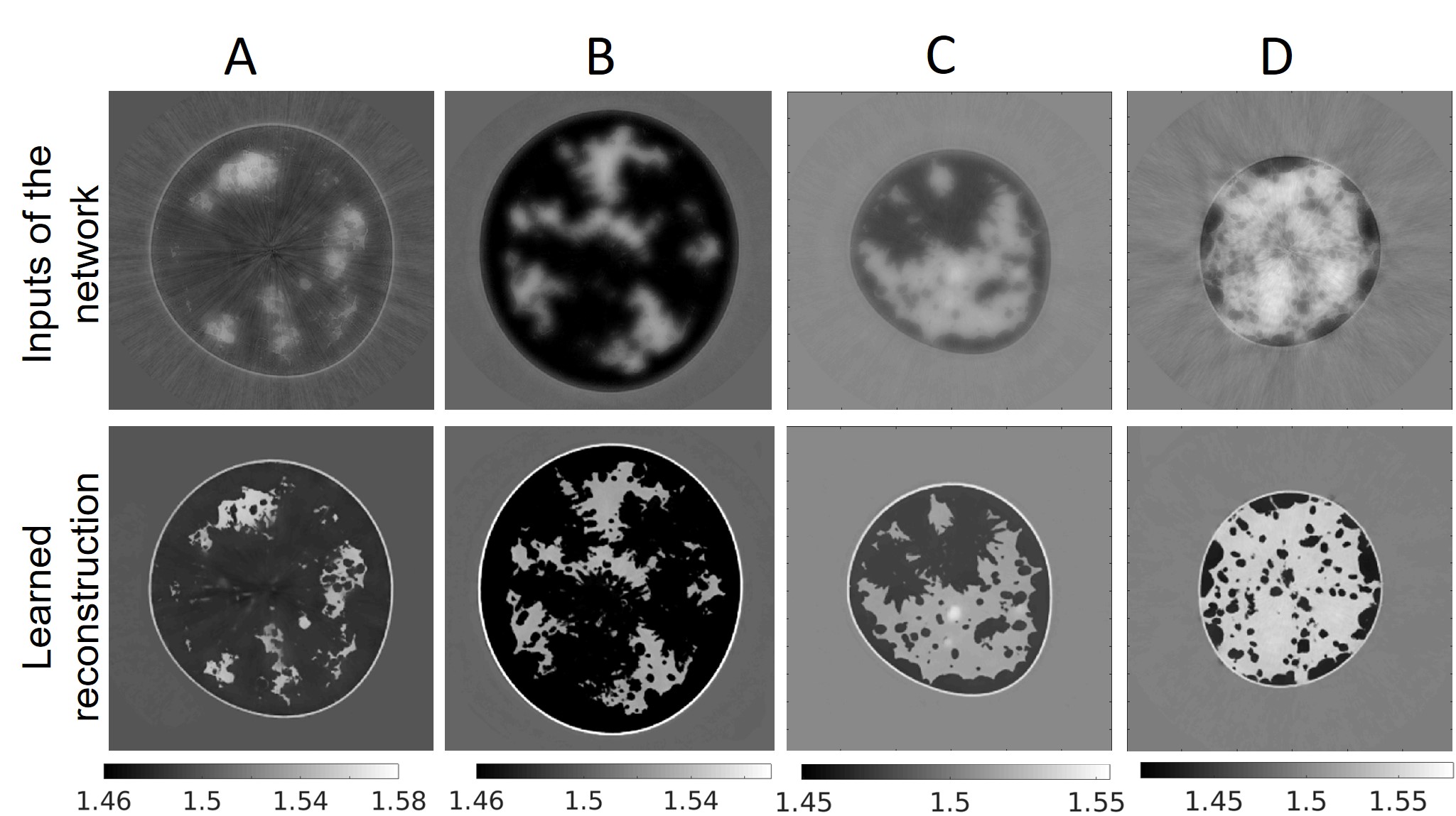}
\caption{Case study 2: Reconstructed SOS images of the phantoms shown in Fig. \ref{fig:ref_recon_examples} using a machine learning-based method. Top row : input to the neural network; Bottom row: the corresponding estimated image. From left to right: breast type A-D. The units are ($mm/\mu s)$.}\label{mapping1}
\end{figure}

\begin{figure}[tb]
\centering
\includegraphics[width=1\linewidth]{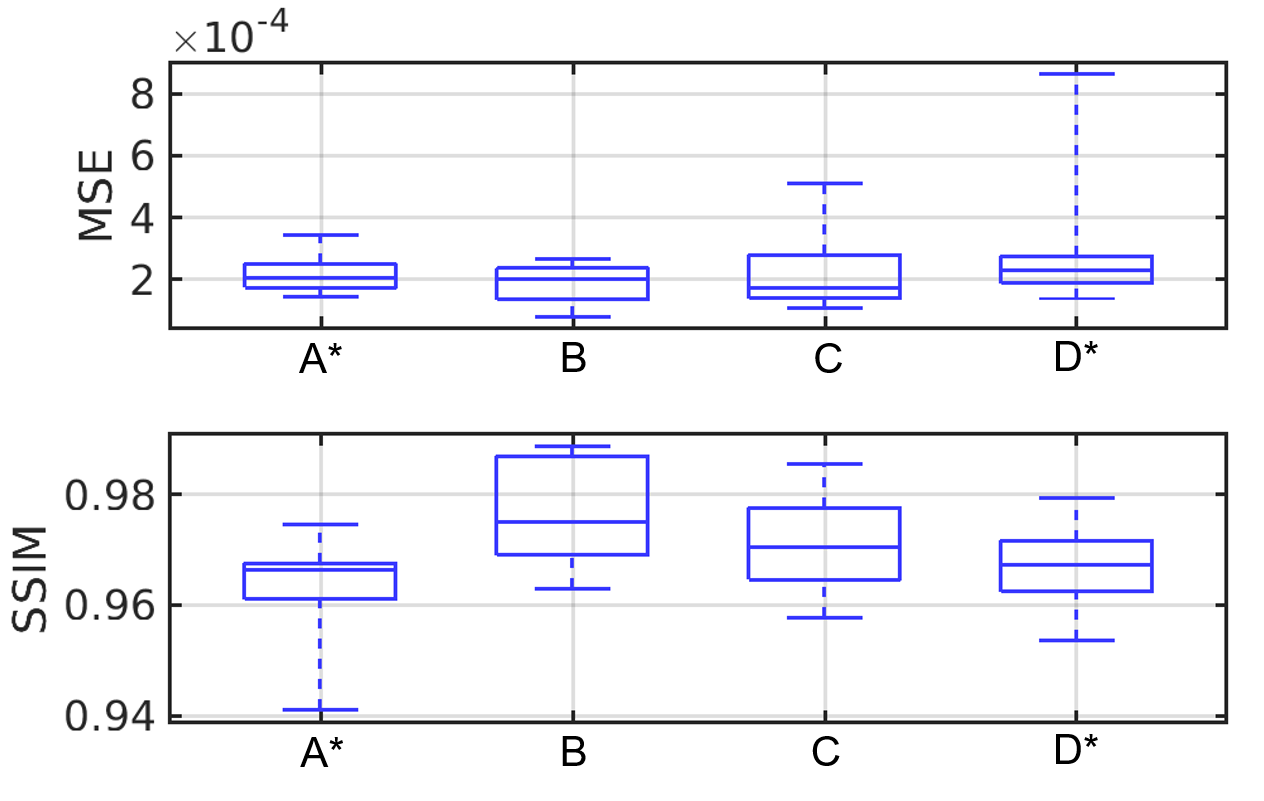}
\caption{Case study 2: Boxplot of MSE and SSIM  value of learned reconstructed results for breast type A-D. The subscript $^*$ denotes out of distribution breast types.}\label{fig:box_dl}
\end{figure}

\begin{figure}[tb]
\centering
\includegraphics[width=0.8\linewidth]{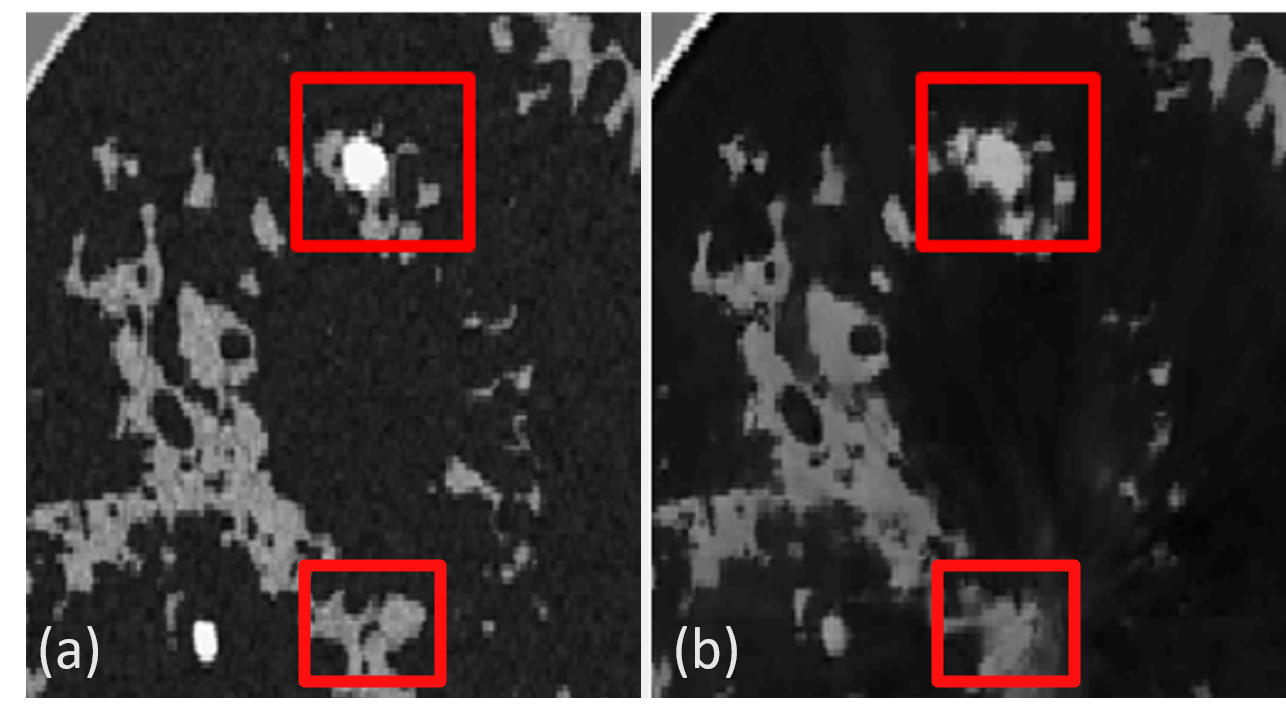}
\caption{Case study 2: False structures
in the reconstructed image. a): The ground truth image. b): The neural network-based reconstructed image that contains small hallucinated features.}\label{fig:hallu}
\end{figure}

Fig. \ref{mapping1} reports examples  of  learned reconstructed images of four breast types. The top row corresponds to the early stopped WISE reconstruction after 35 iterations, and the bottom row shows the output of the neural network. The corresponding ground truth images and reference reconstructions are shown in Fig. \ref{fig:ref_recon_examples}. The proposed learning approach improved the visual  quality of the images, leading to sharper tissue interfaces. 
Table \ref{tab:dl} and Fig. \ref{fig:box_dl} show quantitative evaluations on the test dataset. 
The reported MSE and SSIM values are stratified by breast types: breast types A and D (out of distribution) have a larger median MSE and smaller median SSIM than types B and C (in distribution). While the reported MSE and SSIM are comparable (or sometimes even better) than those reported for the reference reconstructions in Table \ref{tab:ref_recon}, the learned reconstruction method may mistakenly introduce some fine structures (\emph{hallucinations}\footnote{Hallucinations are a specific type of image artifact that are attributable to the prior employed by a reconstruction method and cannot be produced by use of the measurement data alone.}) that are not existing in the ground truth image \cite{9424044}. An example is shown in Fig. \ref{fig:hallu}. This case study demonstrates that, while deep learning methods can be used to enhance perceived and quantitative image quality, their results must be interpreted with particular care due to the possibility of introducing \emph{hallucinated} features in the image. 

\begin{table}[htb]
\caption{Case study 2: mean square error (MSE) and structural similarity index measure (SSIM).}
\label{tab:dl}
\centering
\begin{tabular}{|c|c|c|}
\hline
Breast type & MSE (std) & SSIM (std) \\
\hline
A & 2.165e-04 (8.413e-5) & 0.9675 (0.0081) \\
B & 1.973e-04 (7.9898e-5) &  0.9707 (0.0079)  \\
C & 2.160e-04 (1.149e-4) & 0.9788  (0.0069) \\
D & 2.887e-04 (2.009e-4) & 0.9651 (0.0073) \\
\hline
\end{tabular}
\end{table}

\section{Conclusion}
\label{sec:conclusions}
In this work, procedures were established  by which 3D NBPs can be computed for use in large-scale virtual imaging trials of 2D or 3D breast USCT.
This will, for the first time, permit 3D realistic NBPs to be computed that possess varying shape, acoustic properties, tissue texture, and tumors.
This was accomplished by adapting VICTRE tools to USCT imaging. While some modeling choices and simplifications had to be made, the modular and flexible implementation of the phantom generation procedures allows for additional customizations of the NBPs. For example, future studies may include additional tissue type, use of different lesion models, analyze the detectability of microcalcifications, or develop advanced biomechanical models to capture deformations of the submerged breast due to buoyancy. In summary, the generated NBPs improve the authenticity of USCT virtual imaging studies and can be employed widely for the investigation of advanced image reconstruction methods, objective evaluation of the USCT breast imaging systems, and the development of machine learning-based methods. 



%


\appendices
\section{Publicly released datasets and code availability}
Two datsets have been publicly released on the Harvard Dataverse.

The first dataset \cite{li2021NBPs2D} consists of the 52 slices and corresponding simulated USCT data described in Section \ref{sec:cs_references}. For each slice, the dataset includes tissue label, SOS, AA, and density maps. Each slice contains a variable number of lesions (up to 3) with diameter between 1.5 mm and 5 mm.
The image size is 2560-by-2560 with pixel size at a 0.1 mm resolution. The measurement data have a sampling frequency of 25 MHz and have been perturbed with additive Gaussian white noise as described in \ref{sec:cs_references}.

The second dataset \cite{li2021NBPs3D} consists of four high-resolution 3D NBPs, one for each breast type.
Each NBP contains 3D maps of tissue label, SOS, AA, and density with a resolution of 0.1 mm.  Fig. \ref{3dp} shows 3D rendering of the four public NBPs.

A python package implementing the methods presented in this work is available under GPL-2.0 license from \cite{Fu2021code}.

\section{Supplementary multi-media material}
A video presenting cross-sectional slices of 3D SOS maps (one for each breast type) is included in the supplementary materials.

\section{Computer-simulation of the data acquisition process}
\label{app:k-space-solver}
Pressure wave propagation in the breast tissue was modeled by solving the lossy acoustic wave equation with power law frequency-dependent AA. Specifically, a first-order formulation of the linear acoustic wave equations in heterogeneous media is considered, which is described by the following three coupled differential equations\cite{treeby2010modeling}
\begin{equation}
\left\{
\begin{array}{rl}
  \frac{\partial}{\partial t} \textbf{u}_i =& -\frac{1}{\rho_0} \nabla p_i \\
 \frac{\partial}{\partial t}   \rho_i =& -\rho_0 \nabla \cdot \textbf{u}_i +\int_0^t f_i dt \\
 p =&c_0^2  \left(1-\mu\frac{\partial}{\partial t}(-\nabla^2)^{\frac{y}{2}-1} - \eta(-\nabla^2)^\frac{y-1}{2}\right)\rho_i,
\end{array}
\right.\label{eqn:lossy_wave_eq}
\end{equation}
where $\textbf{u}_i=\textbf{u}_i(\textbf{r}, t), p_i=p_i(\textbf{r}, t), \rho_i =  \rho_i(\textbf{r},t)$ denote the fluctuations of particle velocity, acoustic pressure, and density, respectively, corresponding to the excitation of the $i$-th transducer. The source term $f_i$ has the form  $f_i(\textbf{r}, t) = \delta_{\textbf{r}_i}(\textbf{r})s(t)$, where $\delta_{\textbf{r}_i}$  is the Dirac delta function centered a the location $\textbf{r}_i$ of the $i$-th transducer and $s(t)$ is the pulse profile in eq. \eqref{masource}. The quantities $\rho_0=\rho_0(\textbf{r})$ and  $c_0 = c_0(\textbf{r})$ denote the density and SOS of the medium. The quantities $\mu$ and $\eta$ are defined as
\begin{align}
\mu(\textbf{r}) &= -2\alpha_0(\textbf{r})c_0(\textbf{r})^{y-1}, \\
\eta(\textbf{r}) &= 2\alpha_0(\textbf{r})c_0(\textbf{r})^y \tan \left(\frac{\pi y}{2}\right),
\end{align}
where $\alpha_0(\textbf{r})$ is the AA coefficient and $y$ is the AA exponent. As explained in Section \ref{sec:att_exp}, $y$ was assumed to be spatially homogeneous and its numerical value was determined for each phantom as a function of the fatty and glandular volume fraction. Equations \eqref{eqn:lossy_wave_eq} were discretized on a uniform Cartesian grid and solved using a time explicit pseudospectral k-space method \cite{tabei2002k}. Acoustic transducer locations were approximated by using the center of the pixel to which they belong to. Discretization parameters are reported in Table \ref{tabsystems}. Note that, while finite difference or finite volume discretizations usually require about 10 points per wavelength (ppw), pseudospectral method can correctly capture the solution with as little as 2 ppw. A high-performance GPU-accelerated implementation of the psuedospectral k-space wave solver, developed by the authors\cite{huang2013full,matthews2018parameterized} was employed to perform the simulations. The amplitude of pressure at all transducer locations was recorded as a function of time.

\begin{table}[htb]
\centering
\caption{Imaging system discretization parameters}
\begin{tabular}{| c | c |}
\hline
 Computational grid & 2560 by 2560 pixels\\
 {} &(0.1 mm pixel size, $\sim$6.5 ppw) \\
 \hline
  Time step size & 1/50 $\mu$s, CFL number =0.3  \\
  \hline
  Simulation time  & 170 $\mu$s,  8500 time steps  \\
  \hline
  \end{tabular}
  \label{tabsystems}
\end{table}

\section*{Acknowledgment}
This work was supported in part by NIH awards R01EB028652 and R01NS102213 and NSF award DMS1614305. Computational resources for this work were granted to the authors by the Blue Waters sustained-petascale computing project, which is supported by the National Science Foundation (awards OCI-0725070 and ACI-1238993) the State of Illinois, and as of December, 2019, the National Geospatial-Intelligence Agency. Blue Waters is a joint effort of the University of Illinois at Urbana-Champaign and its National Center for Supercomputing Applications.

\ifCLASSOPTIONcaptionsoff
  \newpage
\fi



%
%
\bibliography{NBP} 
\bibliographystyle{IEEEtran}


%

\begin{IEEEbiography}[{\includegraphics[width=1in,height=1.25in,clip,keepaspectratio]{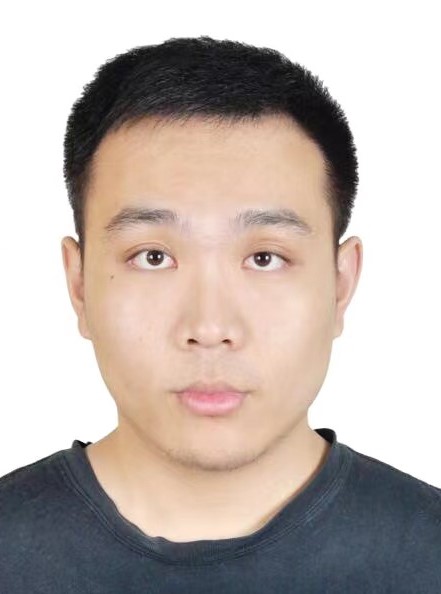}}]{Fu Li}
received the B.S. degree in Information and Computing Science
from Sun Yat-Sen University, Guangzhou, China
in 2016. He is currently pursuing the
Ph.D. degree in bioengineering with the University of Illinois at Urbana–Champaign, Urbana, IL, USA.

His research interests include ultrasound computed tomography and machine learning in medical imaging applications.
\end{IEEEbiography}
\vspace{-1cm}
\begin{IEEEbiography}[{\includegraphics[width=1in,height=1.25in,clip,keepaspectratio]{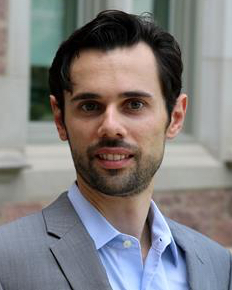}}]{Umberto Villa}
received the B.S. and M.S. degrees in Mathematical Engineering from Politecnico di Milano, Milan, Italy in 2005 and 2007, and the Ph.D. degree in Mathematics from Emory University, Atlanta, GA in 2012. He joined Washington University in St Louis, Missouri, as a Research Assistant Professor of Electrical \& Systems Engineering in 2018. 

Dr. Villa’s research interests lie in the computational and mathematical aspects of large-scale inverse problems, imaging science, and uncertainty quantification. A strong component of his work includes developing scalable efficient algorithms for integrating data (images, experimental measurements or observations) and mathematical models with applications to biomedical and engineering relevant problems.
\end{IEEEbiography}
\vspace{-1cm}
\begin{IEEEbiography}[{\includegraphics[width=1in,height=1.25in,clip,keepaspectratio]{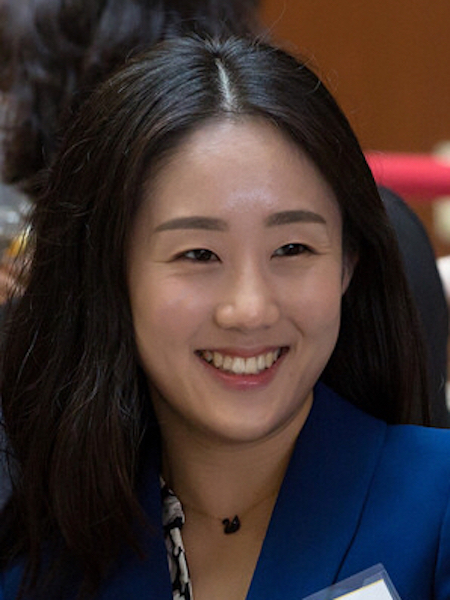}}]{Seonyeong Park}
received the B.S. degree in electronics, computer, and telecommunication engineering and the M.S. degree in information and communications engineering from Pukyong National University, Busan, Korea, in 2011 and 2013, respectively, and the Ph.D. degree in electrical and computer engineering from Virginia Commonwealth University, Richmond, VA, USA, in 2017. She is currently a postdoctoral research associate at the Department of Bioengineering, University of Illinois Urbana-Champaign, Urbana, IL, USA. 

Her research interests include virtual imaging trials, photoacoustic computed tomography, functional imaging, image reconstruction, and inverse problem.
\end{IEEEbiography}
\vspace{-1cm}
\begin{IEEEbiography}[{\includegraphics[width=1in,height=1.25in,clip,keepaspectratio]{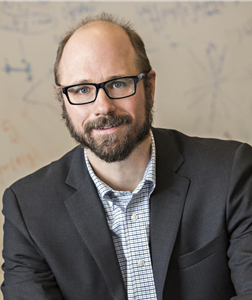}}]{Mark A. Anastasio}(Senior Member, IEEE) earned his Ph.D. from the University of Chicago in 2001. He is currently a Donald Biggar Willett Professor in Engineering and the Head of the Department of Bioengineering at the University of Illinois at Urbana-Champaign.  He is a Fellow of the SPIE, American Institute for Medical and Biological Engineering (AIMBE) and International Academy of Medical and Biological Engineering (IAMBE).  

Dr. Anastasio’s research broadly addresses computational image science, inverse problems in imaging, and machine learning for imaging applications.  He has contributed broadly to emerging biomedical imaging technologies that include photoacoustic computed tomography, ultrasound computed tomography, and X-ray phase-contrast imaging. His work has been supported by numerous NIH grants and he served for two years as the Chair of the NIH EITA Study Section.  
\end{IEEEbiography}

\end{document}